\documentclass{ws-procs9x6}

\begin{document}

\title{Transport Coefficients in Color Superconducting Quark Matter}

\author{Cristina Manuel\footnote{\uppercase{W}ork
supported by \uppercase{MCYT} (\uppercase{S}pain) under grant \uppercase{FPA} 2001-3031.}}

\address{Instituto de F\'{\i}sica Corpuscular\\
Universitat de Val\`encia -C.S.I.C. \\
Edificio Institutos de Paterna,\\ Apt. 2085, 46071 Val\`encia. Spain \\
E-mail: Cristina.Manuel@ific.uv.es}

\maketitle

\abstracts{Transport coefficients of dense quark matter are
needed to study properties of compact stars. They can tell us
about the cooling, vibrational and rotational
properties of the star. We report below a computation of the shear
viscosity in the CFL phase at low temperature $T$. CFL quark matter behaves as
a superfluid, and at low $T$ its transport properties are dominated by the collisions of
superfluid phonons.}

\section{Introduction}

The microscopic understanding of electromagnetic superconductivity
is based on the seminal works of Bardeen, Cooper and Schrieffer. The BCS
theory explains the main phenomenological properties of the superconductors,
such as the Meissner effect, and the appearance of an energy
gap $\Delta$ in the quasiparticle spectrum. It also explains
the fact that superconductors are both good heat conductors and superfluids,
with exponentially suppressed transport coefficients $\sim \exp[- \Delta/T]$
for low temperatures $T \ll \Delta$.

At very high density QCD  is in a color superconducting phase\cite{Rajagopal:2000wf,Alford:1999mk}.
Our
present microscopic knowledge of this phenomena relies on the
BCS techniques adapted to dense quark matter. From first principles one can
evaluate the values of the fermionic gaps and gluon masses.
There are some macroscopic
properties of dense quark matter that have not been yet extensively
studied. Should we expect the transport coefficients in a
color superconductor to behave as in an electromagnetic
superconductor? Naively, one would say so, but the answer is
slightly more elaborated. Quarks have different quantum numbers
 (spin, color, flavor), and the way they
 condense depends on several parameters, such as the quark
masses and chemical potentials. Then one talks about different superconducting
phases (CFL, 2SC, crystalline phases, gapless phases, etc),
which differ basically on the way the quarks pair, and on
the  local or global symmetries that are spontaneously broken.
Because the hydrodynamic regime of a system depends on the low energy
spectrum of the theory, we could simply expect that transport coefficients of
the different superconducting phases to behave very differently.

Let us insist on the relevance of transport phenomena
in astrophysics\cite{Weber:2004kj}. While with the equation of state one can  determine
the mass and radius of a star, with the transport coefficients one
can study its cooling, and its vibrational and rotational properties.
Thus, they are essential for detecting signatures of quark matter
in compact stars in any of its possible phases.

It has been established that the viscosities put stringent tests to astrophysical
models for very rapidly rotating stars, such as for millisecond pulsars.
This is based on the existence of r(otational)-mode instabilities in all
relativistic rotating stars\cite{Andersson:1997xt}, which are only
suppressed by sufficiently large viscosities.
Madsen actually ruled out\cite{Madsen:1999ci}
quark matter in both the CFL and 2SC phases in  millisecond pulsars,
because of their low
viscosities.  This last conclusion, however, was
based on a wrong estimate, as it was assumed a behavior of the
viscosities as in an electromagnetic superconductor
$\sim \exp[-\Delta/T]$, for values of the temperature $T$ of the order of
0.1 MeV and below. These estimates are not correct. In collaboration
with Antonio Dobado and Felipe Llanes-Estrada we have computed the
shear viscosity in the low $T$ regime of the CFL phase\cite{Manuel:2004iv}
that I report below. A previous effort of studying transport properties in the
2SC phase was done in \cite{Litim:2001je}.

\section{Transport in the CFL phase}

Transport properties in the CFL phase of QCD at low $T$ are not dominated by
the quarks, which certainly give a contribution of the
 sort $\sim \exp[-\Delta/T]$.
 In this phase the diquark condensate breaks
spontaneously the baryon symmetry $U_B(1)$, and CFL quark
matter is a superfluid. Chiral symmetry is also spontaneously
broken. Associated to those breakings, there are Goldstone bosons,
which are light degrees of freedom.
 In addition, there is an unbroken $U(1)$ subgroup whose gauge boson,
 a combination of
the photon and one gluon, is massless at zero
temperature and can be viewed as the in-medium photon.  A CFL quark star
is electrically neutral, both for the real and
in-medium electromagnetism, but at finite temperature
one may also expect to find electrons. All
the above mentioned particles are light and their
contribution to the transport coefficients in this phase is bigger
than that of the gapped quarks. Let us mention  that we are considering
a CFL quark star after the deleptonization era,
so that all the neutrinos have escaped from the star.

Chiral symmetry is not an exact symmetry of QCD. Therefore, the
associated (pseudo) Goldstone bosons  are massive. Their masses
are estimated to be in the range of the tens of MeV. At sufficiently
low temperatures, their contribution to the viscosities is Boltzmann suppressed, and
 transport is dominated by the lightest particles.
In particular, the highest contribution\cite{Shovkovy:2002kv} is given by the Goldstone
boson associated to baryon symmetry breaking, the superfluid phonon, which
remains massless. There is also a contribution of the in-medium
electromagnetism, but this turns out to be negligible.

The low energy effective field theory for the superfluid phonon
has been constructed by Son\cite{Son:2002zn}. He realized that the different
coefficients of the corresponding effective Lagrangian can be fixed with the
knowledge of the equation of state of dense quark matter. At high density, the effective
Lagrangian reads
\begin{equation}
\label{L-BGB-0}
 L_{\rm eff}  = \frac{3}{4 \pi^2}
\left[ (\partial_0 \varphi - \mu)^2 - (\partial_i \varphi)^2
\right]^2 \ .
 \end{equation}

The equations of motion associated to $\varphi$ can be interpreted
as hydrodynamic equations for a relativistic superfluid\cite{Son:2002zn}, where the velocity
is given by the gradient of the phase of the condensate, as in Landau's
two fluid model of superfluidity. At finite $T$ they need modifications, as
the different collective modes are thermally excited, and they conform
a different - and dissipative - fluid. At very low $T$, dissipative effects
are dominated by the collisions of the superfluid phonons.

Superfluidity in He$^4$ is understood as a consequence of
Bose-Einstein condensation. There is also a spontaneous breaking
of a global $U(1)$ symmetry, with the corresponding appearance of
a massless collective mode (or Goldstone boson, in the high energy
language). The phonon moves at the speed of sound $v$ of the system.
While in principle He$^4$ and the CFL matter have little in
common, the physics of the collective mode or phonon  in the two
cases exhibits many similarities, and one can then talk about
universality. In particular, we found the same  momentum and $T$
dependence for their damping rates. Also the mean free path has
the same $T$ dependence. Universality aspects of these two systems
are under investigation.

The physics associated to the superfluid phonon is peculiar.
Even if at finite $T$ almost all particles attain a thermal mass, this
is not so for the phonon, as thermal effects do not represent
a violation of the $U_B(1)$ symmetry. The superfluid phonon suffers Landau damping,
which for $p_0, p \ll T$ reads
\begin{equation}
\label{Ldamp-psmall}
 {\rm Im}\, \Pi (p_0,{\bf p})  \approx    \frac{8 \pi^5}{1215}
\frac{T^4}{ v^8 \mu^4} \frac{ p^3_0} { p} \, \Theta(v^2 p^2 -
p^2_0) \ .
\end{equation}
Evaluated on-shell, it gives account of its finite lifetime
in the thermal bath.

Shear viscosity describes the relaxation of the momentum components
perpendicular to the direction of transport, and it is usually
dominated by large-angle collisions. In the CFL fluid it is not the
case. The reason is that the differential cross section of binary collisions
mediated by phonon exchange is divergent for small-angle collisions. It
is only regulated by the finite width of the phonon, which effectively
amounts to a resummation of a certain class of diagrams in the computation. Ultimately,
it turns out that it is more efficient to achieve a large-angle
collision by the addition of many small-angle scatterings.

With Son's effective field theory, it is possible to compute the
mean free path associated to both large-angle and small-angle binary
collisions. The first one behaves\cite{Shovkovy:2002kv}
as $\sim \mu^8/T^9$ and for the values corresponding to
a cold compact star, it exceeds the typical radius of a star $R \sim 10$ km, thus being
irrelevant for transport.
We have computed numerically the mean free path for the second ones,
which parametrically behave as $\sim \mu^4/T^5$. It also exceeds the radius of
a star for $T \ll 0.06$ MeV, the value at which the CFL fluid in the star would
behave as a perfect superfluid.

In Ref.\cite{Manuel:2004iv} we have computed numerically the shear viscosity in the CFL phase within kinetic theory,
taking into account the dominant processes in the collision term. The result
is consistent with a parametric behavior of the sort $\eta \sim \mu^4/T$. This means
that the number of collisions necessary to make a large-angle collision
from many small-angle ones do not depend parametrically on $T$, as one
could naively expect. This is also what happens for superfluid He$^4$, see Ref.\cite{maris}.  

Our result could be improved by taking into account two different 
corrections. Namely, a better determination of the phonon interactions
(determined with a more accurate equation of state of
CFL quark matter),
and the inclusion of the LPM effect in the collision term. We do not expect,
however, that neither of these two corrections would change drastically the
final result of the shear viscosity.

\section{Conclusions}

Viscosities of the different color superconducting phases are needed for
an analysis of r-mode instabillities of compact stars, which tell us about
the maximum stable frequency of the star.
 We have focused our efforts to study
the shear viscosity of the CFL phase, while the bulk viscosities are
also needed and will be the subject of a different project.
 Our results indicate that for $T \ll 0.06$ MeV 
the viscosity  in the phonon fluid is absent and it cannot  damp the r-modes
of millisecond pulsars. However, other sources of dissipation should  be considered 
as well. In particular, in a rotating superfluid there are vortices.
Particles scattering off vortices might contribute substantially to the 
viscosity
of the rotating star. An analysis of the friction of the superfluid and the surrounding crust
might also be relevant for the study of r-modes. So it is premature to discard
the CFL quark matter in a millisecond pulsar, and more work is needed to
draw firm conclusions about this point.
Obviously, it would be equally important to evaluate the viscosities of other
color superconducting phases.

\end{document}